# Zero-Day Threats Detection for Critical Infrastructures

Mike Nkongolo [0000-0003-0938-113X] and Mahmut Tokmak [0000-0003-0632-4308]

University of Pretoria, South-Africa
mike.wankongolo@up.ac.za
Mehmet Akif Ersoy University, mahmuttokmak@mehmetakif.edu.tr

**Abstract.** Technological advancements in various industries, such as network intelligence, vehicle networks, e-commerce, the Internet of Things (IoT), ubiquitous computing, and cloud-based applications, have led to an exponential increase in the volume of information flowing through critical systems. As a result, protecting critical infrastructures from intrusions and security threats has become a paramount concern in the field of intrusion detection systems (IDS). To address this concern, this research paper focuses on the importance of defending critical infrastructures against intrusions and security threats. It proposes a computational framework that incorporates feature selection through fuzzification. The effectiveness and performance of the proposed framework are evaluated using the NSL-KDD and UGRansome datasets in combination with selected machine learning (ML) models. The findings of the study highlight the effectiveness of fuzzy logic and the use of ensemble learning to enhance the performance of ML models. The research identifies Random Forest (RF) and Extreme Gradient Boosting (XGB) as the top performing algorithms to detect zero-day attacks. The results obtained from the implemented computational framework outperform previous methods documented in the IDS literature, reaffirming the significance of safeguarding critical infrastructures from intrusions and security threats.

## 1 Introduction

Recent advances in technologies such as the Industrial Internet of Things (IIoT), wireless systems, and the Web of Things (WoT) have raised some threats that endanger the authenticity, trustworthiness, and cybersecurity of various critical infrastructures. Critical infrastructures are any vital entities (banking, energy, telecommunications, etc.) that are so important to a country's daily operations that their disruption would have disastrous consequences for the country's security and daily operations [1]. Companies can employ an array of countermeasures, such as firewalls, anti-virus programs, anti-malware technologies, and intrusion prevention techniques, to defend themselves against malicious incursions [2]. An Intrusion Detection System (IDS) is acknowledged as one of the most effective tools for securing critical infrastructures from suspicious behavior among all security protocols [3]. This article describes a machine learning (ML) strategy that used fuzzy logic to protect critical infrastructures from unknown network attacks. The following ML techniques were considered: extreme gradient boosting (XGB), extra-trees (ET), decision trees (DT), random forest (RF), naive Bayes (NB), and support vector machines (SVM). The NSL-KDD and UGRansome datasets were also used to assess the performance of the models described in the study [3]. The first one might be thought of as a legacy dataset that contains outmoded properties that have not changed since it was created in 1990. The latter might be viewed as a unique and novel dataset of zero-day threats that was developed in 2021 by Mike Nkongolo et al. [4–6]. A 0-day exploit is an unknown malicious intrusion [7] that can damage critical infrastructures. The penetration occurs without being detected by the IDS. As a result, IDS vendors have zero days to develop keys to recognize them [6]. This makes 0-day intrusion a serious security threat for any critical infrastructure. Consequently, to safeguard critical infrastructures from 0-day exploits, this study employed the UGRansome and NSL-KDD datasets, which underwent a fuzzy-based feature selection process, to develop a computational framework. Furthermore, the implementation of the feature selection (FS) technique is crucial in the proposed computational framework to ensure that only the best (most optimal) patterns are chosen for the modeling process. Moreover, the fuzzy logic's selection criteria used the ET classifier. In the experiment, four feature vectors were produced by the fuzzy-based feature selector for the multi-class and binary classification processes. The modeling technique made use of the four attribute vectors. Using those four specific features, this study evaluated the models' performance and contrasted the findings with those of existing ML methods. The experimentation proved that the classifiers under consideration work better when fuzzy logic is used for FS. The rest of this article is divided into the following sections: The related work is described in Section 2; experimental datasets are outlined in Section 3; and the background of the ML models are introduced in Section 4. Sections 5 outlines the proposed computational framework. An explanation of the findings is included in Section 6 which also includes experimental setups. This study report is concluded in Section 7.

## 2   Related Works

The NSL-KDD dataset was utilized by Zhang et al. [8] to develop a deep learning-based IDS architecture. In the study, the authors developed an auto-encoder technique to choose the most pertinent features that the deep neural network (DNN) used for classification. The auto-encoder-based DNN model's performance was evaluated using the accuracy, recall, precision, and F1 score. The findings showed that the auto-encoder enhanced the DNN classification performance with 79.74% accuracy, 82.22% precision, 79.74% recall, and 76.47% F1-score. To increase the detection accuracy, the authors may have added extra parameters to the DNN architecture. Tama et al. [9] provided a two-step methodology for intrusion detection systems (TSE-IDS). The TSE-IDS uses a variety of FS techniques in the first stage, including particle swarm optimization (PSO), ant colony algorithm (ACO), and genetic algorithm (GA). Based on the results from the pruning tree (REPT) model, the fitness criterion was able to extract a feature set from the original data. The bagging technique was used as part of an ensemble learning scheme in the second step. The UNSWNB15 and NSL-KDD datasets were used in the experimentation. Moreover, the binary classification configuration was taken into account. Some of the metrics, including precision, accuracy, and sensitivity, were used to assess the effectiveness of their methods. The tests showed that, for the NSL-KDD dataset, TSE-IDS achieved 85.797% accuracy, 88.00% precision, and 86.80% sensitivity. Using the UNSW-NB15 dataset, the TSE-IDS achieved accuracy, sensitivity, and precision values of 91.27%. However, the TSE-IDS was not tested with a multi-class configuration. A comparison of the SVM with rule-based classifiers was performed by Sarumi et al. [10]. A filter and wrapper-based FS approach was used for each of those models. The NSL-KDD and UNSW-NB15 datasets were utilized to evaluate the Apriori and SVM's performance. By concentrating on the NSLKDD results, it was shown that a filter-SVM could achieve 77.17% accuracy, 95.38% recall, and 66.34% precision. The accuracy, recall, and precision of the filter-Apriori approach were 67.00%, 57.89%, and 85.77% respectively. In contrast, the wrapper-SVM obtained a precision of 68%, an accuracy of 79.65%, and a recall of 98.02%. A precision of 85%, an accuracy of 68.6%, and a recall of 58.81% were achieved using the wrapper-Apriori. Most of the research articles discussed in the IDS literature used the legacy NSL-KDD dataset. This research argues that NSL-KDD data does not include 0-day attack properties due to missing patterns to address current network concerns. In addition, this dataset presents different categories of malicious patterns that have changed and evolved. Similarly, the UNSW-NB15 dataset is not strongly correlated to realistic network behaviors and cannot be used to study unknown malware. However, Suthar et al. [4] introduced a novel dataset named Emotet and recommended UGRansome for ransomware detection. On another front, Maglaras et al. [11] reported the UGRansome performance in a cloud-based optimization setting for 0-day threats recognition to protect critical infrastructures from malicious intrusions [12]. More recently, Tokmak [6] compared the performance of the deep forest model with SVM, NB, RF, and DNN algorithms on the UGRansome and NSL-KDD datasets. The range of the obtained results was between 87% to 99%. In the experiments, the UGRansome proved its efficacy in categorizing three predictive classes of 0-day concerns such as anomaly (A), signature (S), and synthetic signature (SS). In addition, the deep forest achieved more successful results with the following algorithms: DNN (97%), RF (97%), and SVM (96%). As such, our research agrees with Shankar et al. [5] and Rege and Bleiman [1] who have recently recommended UGRansome to be used in the recognition of unknown networks attacks like ransomware and zero-day attacks.

## 3    Experimental Datasets

The NSL-KDD dataset is utilized in this study to evaluate the effectiveness of the suggested fuzzy-based computational model [2]. It is a popular corpus in the IDS field compared to the UGRansome dataset. It has served as the foundation for numerous ML-based IDS studies. R2L, normal, U2R, Probe, and DoS traffic trace categories are all included in the NSL-KDD corpus [2]. The NSL-KDD has two subsets: the NSL-KDD Train and the NSL-KDDTest+ [2]. The NSLKDD was split into the NSL-KDD-Train+ and NSL-KDD-Val. Furthermore, 20% of the overall NSL-KDD-Train set is made up of the NSL-KDD-Val which has 80% of the original training data. The chosen models were computed on the NSL-KDD-Train+ partition, the trained models were evaluated using the NSL-KDD-Val, and the technique was tested using the NSL-KDD-Test+. The validation stage ensures that the algorithms used are not prone to overfitting ML models [2]. The features that make up the NSL-KDD attributes are shown in Table 1. Table 2 presents the distribution of the values for each threat category for each subset. In this study, the NSL-KDD is compared to the UGRansome dataset [3,4,12]. However, the proposed fuzzy-based framework's performance was evaluated using the UGRansome data.

Table 1: A description of the features in the NSL-KDD dataset

| No | Name | No | Name |
|-----|----------------|------|---------------|
| f1 | Protocol | f2 | Service |
| f3 | Flag | f4 | Duration |
| f5 | Bytes | f6 | Error rate |
| f7 | Urgent | f8 | Hot |
| f9 | Failed logins | f10 | Dst count |
| f11 | Logins | f12 | Dst srv count |
| f13 | Num compromised | f14 | Root shell |
| f15 | Su attempted | f16 | Num root |
| f17 | Num shell | f18 | Access file |
| f19 | Outbound cmds | f20 | Host login |

The UGRansome dataset is a comprehensive anomaly detection dataset that encompasses multiple attack categories, including Signature (S), Anomaly (A), and Synthetic Signature (SS). In addition, the dataset was divided into two partitions, namely, the UGRansome19Train and UGRansome-Val. The UGRansomeVal represents 30% of the entire UGRansome19Train which constitutes 70% of the original training data. The selected ML models were computed on the UGRansome19Train partition, the validation set was utilized to evaluate the trained models while the UGRansome18Test was mainly used for testing purposes. The UGRansome validation step guarantees that the ML classifiers used in the experiments are not prone to overfitting.

Table 2: Distribution of subsets in the NSL-KDD dataset

| Dataset | Normal | DoS | Probe | R2L | U2R | Total |
|---|---|---|---|---|---|---|
| KDDTrain | 57,343 | 35,927 | 10,656 | 895 | 42 | 115,973 |
| KDDTrain+ | 40,494 | 24,478 | 7,717 | 649 | 42 | 84,480 |
| KDDVal | 15,849 | 10,449 | 1,939 | 146 | 10 | 21,493 |
| KDDTest+ Full | 8,711 | 6,458 | 1,754 | 1,421 | 100 | 22,544 |

The dataset includes three categories and their respective labels (S, SS, and A). Each predictive class has been labeled to include 0-day threats such as Locky, CryptoLocker, advanced persistent threats (APT), SamSam, and Globe [3]. Table 3 presents the UGRansome design as it was used to test and train the fuzzy based FS model [3]. The values distribution per attack categories for each data subset is outlined in Table 4.

Table 3: A description of the features in the UGRansome dataset

| No | Name | Type | No | Name | Type |
|---|---|---|---|---|---|
| f1 | SS | Categorical | f2 | Cluster | Numeric |
| f3 | S | Categorical | f4 | A | Categorical |
| f5 | Spam | Categorical | f6 | BTC | Numeric |
| f7 | Blacklist | Categorical | f8 | Bytes | Numeric |
| f9 | Nerisbonet | Categorical | f10 | USD | Numeric |
| f11 | UDP scan | Categorical | f12 | JigSaw | Categorical |
| f13 | SSH | Categorical | f14 | Port | Numeric |
| f15 | DoS | Categorical | f16 | CryptoLocker | Categorical |
| f17 | Port scanning | Categorical | f18 | WannaCry | Categorical |

Table 4: Distribution of subsets in the UGRansome dataset

| Dataset | A | S | SS |
|---|---|---|---|
| UGRansome19Train | 40,323 | 25,822 | 9,656 |
| UGRansomeVal | 11,869 | 9,439 | 1,736 |
| UGRansome18Test | 4,701 | 3,408 | 6,954 |
| Total | 56,893 | 38,669 | 18,346 |
| Average | 19,964 | 12,889 | 6,115 |

## 4    Machine Learning Background

A DT is a hierarchical structure that uses a series of if-else conditions to make decisions or predictions [2]. The mathematical formulation involves defining the splitting criteria and calculating impurity measures such as Gini Index or Entropy.

1. **Splitting Criteria**: Let D be the dataset at a particular node of the DT, and X be the feature set. The splitting criteria aim to find the best feature and threshold value that maximizes the separation of classes or reduces impurity [2]. For a binary split, the splitting criterion can be defined as:

$$Splitting_{criterion}(D,X) = argmax_{Gain}(D,X) \tag{1}$$

   Where Gain(D, X) represents the gain obtained by splitting the dataset D based on feature X.

2. **Entropy**: The Entropy (H) is computed as follows:

$$H(D) = -\sum p_i \times log(p_i)$$

In this study, a DT has been used to address the UGRansome multi-class classification problem, as shown in Fig. 1. The root node, intermediate node, and leaf node of the UGRansome's data structure have been demonstrated.

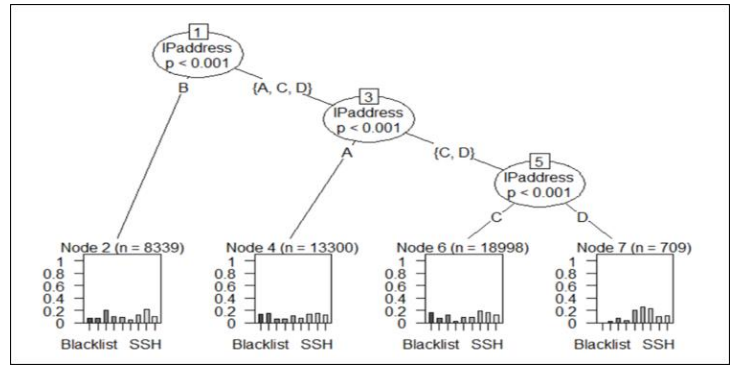

Fig.1: DT of the UGRansome dataset

Fig. 1 involves the use of a DT to classify 0-day threats based on IP addresses of the UGRansome dataset [3,12]. The root node represents the IP address, which is divided into four classes (A, B, C, and D) based on the presence of malicious activity [3]. The tree is recursively split at intermediate nodes until the final prediction (Blacklist and SSH attacks) is reached. The leaf nodes represent the final predictions made by the DT algorithm.

## 4.1    Tree-based ensemble classifiers

The tree-based ensemble classifiers XGBoost, ET, and RF are selected and used in this study. DTs are the building blocks utilized in these algorithms.

The RF is an ensemble learning algorithm that combines multiple DTs to make predictions [12]. Each DT in the RF is built on a different subset of the training data and uses a random selection of features. The final prediction is generally made by aggregating the predictions of all the individual trees [3]. The RF algorithm leverages the diversity and collective decision-making of multiple DTs to improve the overall prediction accuracy and generalization capability [2]. Extra Trees (ET), or extremely randomized trees, further randomizes the DT splits compared to traditional DTs. The mathematical formulation is similar to the DT, but the splitting criteria involve additional randomization [2,6]. In turn, XGBoost is an ensemble-based ML technique that utilizes multiple DTs [2] by combining the outputs of weak learners through gradient boosting. The implementation involves the following objective function and the update rule:

1. **Objective Function**:

$$Obj(\theta) = L(y,y_*) + \gamma(f),\qquad(2)$$

Where L is the loss function, y is the true labels, $y_*$ is the predicted labels, $\gamma(f)$ is the regularization term for the model complexity, and $\theta$ represents the model parameters.

2. **Update Rule**:

$$f_t = f_{(t-1)} + \psi \times h_{t(x)},\qquad(3)$$

where $f_t$ is the prediction at iteration t, $\psi$ is the learning rate, $h_{t(x)}$ is the weak learner at iteration t, and x is the input data.

## 4.2    Naive Bayes

Naive Bayes (NB) is a probabilistic classification algorithm that assumes independence between features given the class variable [3]. The mathematical formulation involves calculating the posterior probability of each class given the input features. Given a dataset D with N samples and K classes, and an input feature vector $X = (x_1, x_2, ..., x_m)$, the mathematical formulation of NB can be expressed as follows:

1. **Prior Probability**:

$$P(C_k) = \frac{N_k}{N}$$

(4)

where $P(C_k)$ represents the prior probability of class $C_k$, $N_k$ is the number of samples in class $C_k$, and N is the total number of samples.

2. **Likelihood**:

$$P(X|C_k) = P(x_1|C_k) \times P(x_2|C_k) \times ... \times P(x_m|C_k)$$

(5)

where $P(X|C_k)$ represents the likelihood of observing the feature vector X given class $C_k$.

### 4.3   Support vector machine

Support vector machine (SVM) is a supervised ML algorithm that can be used for both classification and regression tasks [3]. In the case of binary classification, the mathematical formulation of SVM involves finding an optimal hyperplane that separates the data into two classes while maximizing the margin between the classes [12]. The formulation can be represented as follows: Given a dataset $D = (x_1, y_1), (x_2, y_2), ..., (x_N, y_N)$, where $x_i$ represents the input feature vector of dimension d, and $y_i$ represents the corresponding binary class label (-1 or +1), the objective function of the SVM formulation aims to find a hyperplane characterized by the weight vector w and bias term b, which separates the data points with maximum margin. This function can be formulated as:

$$minimize : (\frac{1}{2}) \times ||w||^2 + C \times \sum (max(0, 1 - y_i \times (w^T \times x_i + b)))$$

(6)

where $||w||^2$ represents the L2-norm of the weight vector w, C is a regularization parameter that controls the trade-off between maximizing the margin and minimizing classification errors, and the second term represents the hinge loss function that penalizes misclassifications [2]. The SVM formulation includes the following constraints:

$$y_i \times (w^T \times x_i + b) >= 1 - \epsilon_i, \forall\ i = 1, 2, ... N$$

(7)

$\epsilon_i >= 0$, for i = 1, 2, ..., N; where $\epsilon_i$ represents the slack variable that allows for the presence of some misclassified or margin-violating samples. The decision function for classifying new input data x can be defined as:

$$f(x) = sign(\sum (\alpha_i \times y_i \times K(x_i, x)) + b)$$

(8)

where $K(x_i,x)$ represents the kernel function that computes the similarity or dot product between the support vectors $x_i$ and the new input data x. The final SVM model is determined by the values of the weight vector w, the bias term b, and the support vectors obtained during the training process.

## 5    The Proposed Fuzzy-based Framework

The proposed method for fuzzy logic-based feature selection (FL-based FS) involves the following steps:

1. **Define the Universe of Discourse**: Let U be the universe of discourse representing the set of all possible features. Each feature x belongs to U, i.e., $x \in U$.

2. **Define Fuzzy Sets**: Identify the fuzzy sets that represent different degrees of relevance or importance of features. Let $A_i$ be a fuzzy set associated with feature $x_i$, where i = 1, 2, ..., N. $A_i$ is defined by a membership function $\mu_i(x_i)$, which assigns a degree of membership between 0 and 1 to each feature $x_i$.

3. **Assess the Relevance or Importance of Features**: Determine the degree of relevance or importance of each feature based on a specific criterion or objective [13]. This can be done by evaluating the membership function $\mu_{i(xi)}$ for each feature $x_i$. The membership function may consider various factors, such as statistical measures to determine the degree of relevance [13].

4. **Rank Features**: Once the degrees of relevance or importance are obtained for all features, the algorithm ranks them in descending order based on their membership values [13]. The ranking indicates the relative significance of each feature in contributing to the desired criterion or objective [13].

5. **Select the Subset of Features**: Choose a threshold or a predetermined number of features to select from the ranked list. Features with higher membership values above the threshold are considered more relevant and selected for further analysis or modeling.

6. **Perform ML**: Use the selected subset of features as input variables for the ML algorithms. The FS process aims to improve the model's performance by reducing dimensionality and focusing on the most informative features [7].

The proposed FL-based FS has been illustrated in Algorithm 1. Moreover, the FL was computed in various stages to select the list of pertinent features ($V = v_1,v_2,v_3,v_4;g_1,g_2,g_3,g_4$) used in the classification process. Table 5 gives information on the selected feature vector and the length of each vector.

---

**Algorithm 1** Fuzzy Logic-based Feature Selection

---

**Require:**
1: Input features $X = \{x_1, x_2, ..., x_n\}$
2: Triangular membership function parameters $a,b,c$ **Ensure:**
3: Selected features $X_{\text{selected}}$
4: **function** FuzzyFeatureSelection($X,a,b,c$)
5:         Normalize input features $X$
6:            Calculate membership degrees $\mu(x;a,b,c)$ for each feature $x \in X$
7:            Initialize feature importance scores $I = \{0,0,...,0\}$
8:        **for** $i = 1$ to $n$ **do**
9:           **for** $j = 1$ to $n$ **do**
10:                $I[i] \leftarrow I[i] + \mu(x_j;a,b,c)$
11:        Sort features based on importance scores in descending order
12:        $X_{\text{selected}} \leftarrow$ top-ranked features from $X$ based on $I$
13:        **return** $X_{\text{selected}}$

---

The approach utilized a triangular membership function characterized by three parameters, namely a, b, and c. This membership function assigned a value between 0 and 1 based on how close the data point is to the center (b) of the triangular function. The general form of a triangular membership function ($\mu$(x)) is:

$$\mu(x) = \begin{cases} \frac{b-a}{x-a} & \text{if } a \leq x \leq b, \\ \frac{c-x}{c-b} & \text{if } b \leq x \leq c, \\ 0 & \text{otherwise}, \end{cases} \tag{9}$$

where $\mu$ represents the membership value for a data point x in the UGRansome fuzzy set. The proposed framework consists of components for data preparation, feature selection, modeling, and evaluation. The numerical/categorical inputs of the experimental datasets are normalized (scaled) during the data preparation/pre-processing phase. The Min-Max scaling function shown in Equation 10 was computed [2].

$$v_{scaled} = \frac{v_n - min(v_n)}{max(v_n) - min(v_n)} \tag{10}$$

This procedure makes sure that the input data stays between [0, 1].

**Performance metrics**

The evaluation metrics employed to assess the models developed in this study include precision (PR), recall (RC), accuracy (AC), F1-score (F1S), and empirical error:

$$PR = \frac{TP}{TP + FP} \tag{11}$$

$$RC = \frac{TP}{TP + FN} \tag{12}$$

$$AC = \frac{TP + TN}{TP + TN + FP + FN} \tag{13}$$

$$F1S = 2 \times \frac{PR \times RC}{PR + RC} \tag{14}$$

$$Error = \frac{FP + FN}{TP + TN + FP + FN} \tag{15}$$

Those metrics were calculated using the Confusion Matrix (CM) [5]: CM:

$$CM = \begin{bmatrix} TN & FP \\ FN & TP \end{bmatrix}$$

There are four key categories to consider in the matrix: True Negative (TN) refers to the accurate identification of regular network patterns as non-malicious, while True Positive (TP) represents the correct identification of intrusions as malicious activities [3]. On the other hand, False Positive (FP) items are nonintrusive activities that are mistakenly categorized as intrusions [12]. Lastly, False Negative (FN) encompasses hostile network patterns that are incorrectly labeled as normal or non-malicious [1]. The F1S is the harmonic mean of the PR and RC. Additionally, the receiver operating characteristic (ROC) curves for each selected method were also computed. The ROC curve was used to calculate the area under the curve (AUC) of each model. The AUC's range is restricted to [0, 1], and values closer to 1 indicate a successful classification method.

## 6    Results

The experimentation was carried out using two classification procedures such as the binary and multi-class classification processes. Five labels from the NSLKDD dataset were taken into consideration during both processes (DoS: 4, Probe: 3, U2R: 2, R2L: 1, and Normal: 0). For the UGRansome dataset, three labels were taken into consideration (SS: 2, S: 1, and A: 0).

### 6.1 Binary classification results

The RF algorithm, which uses $v_2$–$v_4$ and has a validation accuracy (VAC) value of 99.8%, has the highest performance values in terms of VAC. The DT had an accuracy rate of 89.1% and had the highest testing accuracy (TAC) values. The DT used $v_3$ to achieve a VAC of 99.8%, PR of 95.0%, RC of 80.6%, and F1S of 87.0%. The classifier that produced the highest performance results in terms of PR was the RF, which used $v_1$ –$v_4$ to achieve a VAC rating of 99.9%. Similar to this, the best NB model achieved a score of 80.6% in terms of TAC. When compared to the NSL-KDD dataset, the experiment shows a modest performance improvement with the UGRansome dataset. The ROC curves produced by each model (XGB, ET, RF, SVM, DT, and NB) were displayed to assess the effectiveness of the classification process using the AUC metric. The feature vectors $v_1$, $v_3$, and $v_4$ produced similar outcomes. The baseline classifiers (NB and SVM) on the other hand, obtained AUCs of 62% and 64%, respectively. In conclusion, tree-based models had the best binary classification performance. The AUCs measurements found in Fig. 2 provide evidence in favor of this performance. The plot on the left side shows the ROC curves of each model tested and trained with the NSL-KDD. The plot on the right shows the ROC curves of each model tested and trained with the UGRansome.

### 6.2 Multi-class classification results

The XGB had the highest TAC value, with a score of 89.8% using $g_2$ (Table 6). Moreover, the classifier that produced the highest performance values in terms of VAC was the RF, which used $g_1$ –$g_4$ and had a VAC value of 99% (Table 6). The CM depicts that selected models were successful in identifying attack types such as anomaly (A), DoS, probe, normal, signature (S), and synthetic signature (SS) in both datasets (Fig. 3). Fig. 3 portrays the overall classification results. A comparative analysis with existing methodologies of the IDS literature has been illustrated in Table 7.

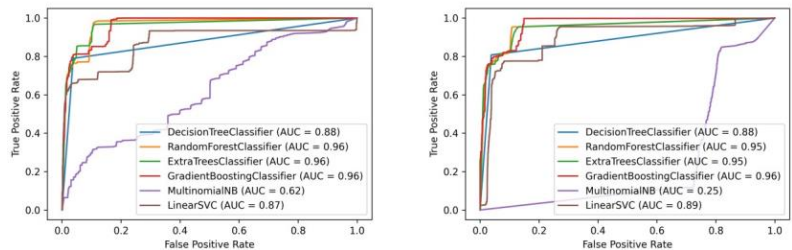

Fig.2: ROC curve analysis of each model using extracted features

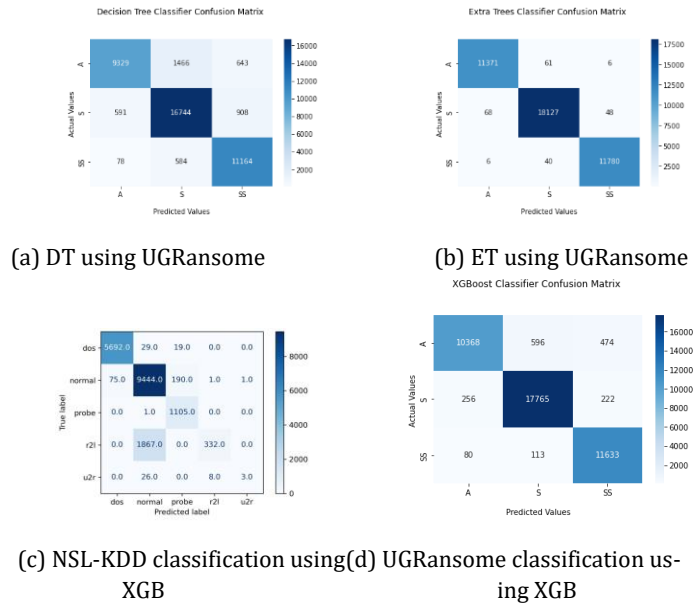

(a) DT using UGRansome

(b) ET using UGRansome

(c) NSL-KDD classification using XGB

(d) UGRansome classification using XGB

Fig.3: Classification results for the UGRansome and NSL-KDD

The proposed methodology outperformed works presented in Table 7 due to the fuzzy logic that acted as a feature selector approach. The proposed methodology surpassed results obtained by Zhang et al. [8], Tama et al. [9], Sarumi et al. [10], Tokmak [6], and Lobato et al. [14].

Table 5: Selected features using the FL procedure. These features/patterns are found in both datasets in detecting zero-day attacks.

| Dataset | Vector | Length | List of features | 0-day detection? |
|---------|--------|--------|------------------|------------------|
| **NSL-KDD** | $v_1$ | 11 | logged in, flag | Yes |
| | | | num access file, num outbound cmds | Yes |
| | $v_2$ | 9 | dst host srv diff host rate | Yes |
| | | | src bytes | Yes |
| | $v_3$ | 9 | num access files, is guest login | Yes |
| | | | same srv rate | No |
| | $v_4$ | 10 | protocol type, flag | Yes |
| | | | land | No |
| | | | same srv rate | No |
| **UGRansome** | $v_1$ | 11 | Locky, CryptXXX | Yes |
| | | | CryptoLocker, EDA2 | Yes |
| | | | Globe | Yes |
| | $v_2$ | 9 | S | Yes |
| | | | B, C, D | No |
| | $v_3$ | 20 | Blacklisting | Yes |
| | | | 1GZKujBR, 8e372GN, Crypt | Yes |
| | | | WannaCry, CryptoLocker, EDA2 | Yes |
| | | | Globe | Yes |
| | $v_4$ | 14 | UDP | No |
| | | | TCP | No |
| | | | B | No |
| | | | Blacklisting, SSH | Yes |
| **NSL-KDD** | $g_1$ | 14 | duration, service, urgent, files | Yes |
| | | | access, outbound | Yes |
| | $g_2$ | 19 | attempted, root, shells, access files count, srv count, error rate, | Yes |
| | | | src bytes | Yes |
| | $g_3$ | 20 | rate, rv rate, host count | Yes |
| | | | same srv rate | No |
| | $g_4$ | 10 | protocol type, flag | Yes |
| | | | land | No |
| | | | same srv rate | No |
| **UGRansome** | $g_1$ | 13 | 1DA11mPS, 1NKi9AK5 | Yes |
| | | | 1GZKujBR, 1AEoiHY2 | Yes |
| | | | Globe | Yes |
| | $g_2$ | 9 | S | Yes |
| | | | 17dcMo4V, 1KZKcvxn | Yes |
| | $g_3$ | 20 | 1BonuSr7, 1sYSTEMQ | Yes |
| | | | 18e372GN, 1Clag5cd, 1DiCeTjB | Yes |
| | | | 1Lc7xTpP, WannaCry, CryptoLocker, EDA2 | Yes |
| | | | NoobCrypt | Yes |
| | $g_4$ | 14 | DMALocker | Yes |
| | | | NoobCrypt | Yes |
| | | | CryptoLocker20 | Yes |
| | | | Blacklisting, SSH | Yes |

Table 6: Binary classification performance results obtained from the UGRansome dataset

| Classifier | Vector | VAC | TAC | PR | RC | F1S |
|---|---|---|---|---|---|---|
| NB | $v_1$ | 84.2% | 78.6% | 77.4% | 61.5% | 74.7% |
| XGB | $v_1$ | 98.3% | 86.0% | 95.1% | 75.2% | 86.0% |
| ET | $v_1$ | 98.6% | 85.4% | 96.0% | 77.0% | 85.6% |
| RF | $v_1$ | 99.9% | 87.5% | 96.5% | 78.5% | 86.4% |
| SVM | $v_1$ | 90% | 80% | 94% | 63% | 75% |
| DT | $v_1$ | **99.3%** | 86.8% | 96.1% | 78.6% | 85.3% |
| NB | $v_2$ | **90.5%** | **80.2%** | 95.3% | 63.6% | 74.7% |
| XGB | $v_2$ | 99.2% | 86.3% | 96.1% | 75.3% | 84.4% |
| ET | $v_2$ | 99.0% | 86.1% | 95.3% | 74.4% | 84.8% |
| RF | $v_2$ | **99.9%** | 87.4% | 96.3% | 76.7% | 85.1% |
| SVM | $v_2$ | 91.1% | **88.8%** | 89.1% | 77.8% | 84.1% |
| DT | $v_2$ | 99.1% | 88.4% | 95.3% | **80.4%** | 87.8% |
| NB | $v_3$ | 88.9% | **80.4%** | 93.1% | 61.4% | 73.3% |
| XGB | $v_3$ | 99.7% | 86.1% | 96.0% | 74.1% | 82.6% |
| ET | $v_3$ | 99.6% | 86.4% | 95.3% | 73.4% | 82.6% |
| RF | $v_3$ | **99.9%** | 87.4% | 97.4% | 76.6% | 85.8% |
| SVM | $v_3$ | 95.1% | 80.4% | 90.1% | 65.2% | 76.4% |
| DT | $v_3$ | **99.9%** | **90.1%** | 96.3% | 85.9% | 88.7% |
| NB | $v_4$ | 88.3% | 80.6% | 98.1% | 60.8% | 74.1% |

| | | | | | |
|---|---|---|---|---|---|
| XGB | $v_4$ | 99.5% | 87.3% 96.3% | 77.5% | 86.3% |
| ET | $v_4$ | 99.8% | 88.8% 96.5% | 78.4% | 86.6% |
| RF | $v_4$ | **99.9%** | 87.3% 97.6% | 76.4% | 85.3% |
| SVM | $v_4$ | **96%** | 81% 91% | 66% | 77% |
| DT | $v_4$ | 99% | **89.0%** 95.8% | 80.7% | 87.1% |

Table 7: Comparative analysis with previous studies

| No | Author | Method | Accuracy | Advantage | Limitation |
|---|---|---|---|---|---|
| 1 | Zhang et al. [8] | DNN | 82% | True positive rate | Biases |
| 2 | Tama et al. [9] | FS | 91% | True negative rate | Data normalization |
| 3 | Sarumi et al. [10] | SVM | 98% | Classification | Empirical error |
| 4 | Suthar et al. [4] | FS | - | Internet protocol | Weak evaluation |
| 5 | Tokmak [6] | DL | 97% | PSO | Feature scaling |
| 6 | Kasongo [2] | GA | 98% | FS | Legacy data |
| 7 | Lobato et al. [14] | SVM | 90% | 0-day detection | False positive/negative |

## 6.3    Zero-day attack prediction

The proposed framework exhibited promising performance in predicting advanced persistent threats (APTs), including the notorious Razy attack (Fig. 4). However, it did not demonstrate its effectiveness in effectively countering such threats, which are known for their sophisticated and evolving nature [3,12]. This discussion underscores the significance of detecting unknown attacks, such as EDA (Fig. 4). To enhance its capabilities, the proposed framework should be expanded to include real-time monitoring and adaptive defense mechanisms, enabling proactive protection against zero-day attacks [15]. In future work, the authors intend to explore the application of a heuristic feature selection technique specifically for zero-day threat detection.

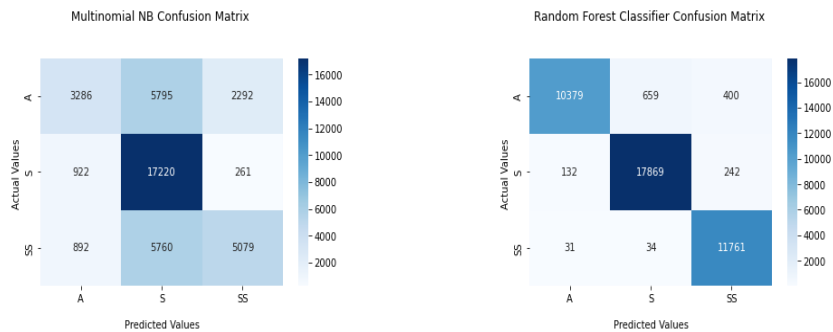

(a) NB using UGRansome             (b) RF using UGRansome

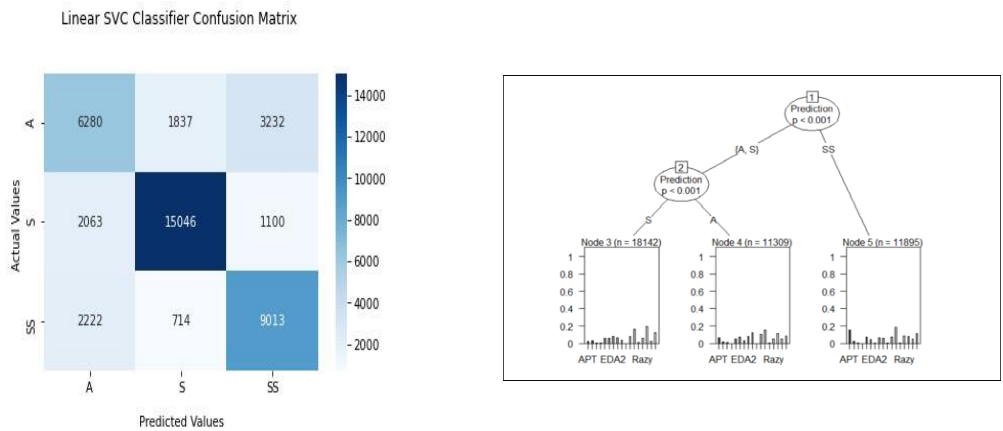

(c) SVM using UGRansome     (d) Zero-day attack prediction using DT

Fig.4: Classification and prediction using the UGRansome

## 7 Conclusion

This paper presents a computational framework for defending critical infrastructures against intrusions and security threats. It incorporates feature selection through fuzzification and evaluates the effectiveness using machine learning models with the NSL-KDD and UGRansome datasets. The study demonstrates the superior performance of fuzzy logic and ensemble learning in detecting zero-day attacks, outperforming previous methods in intrusion detection systems.